\documentclass[aps,prl,twocolumn,groupedaddress,showpacs]{revtex4}
\usepackage{epsfig}

\def\8{\infty}
\def\oh{\frac{1}{2}}

\def\d{\partial}

\def\undertext#1{\vtop{\hbox{#1}\kern 1pt \hrule}}

\def\be{\begin{equation}}
\def\ee{\end{equation}}
\def\bea{\begin{eqnarray} & &}
\def\eea{\end{eqnarray}}

\def\rf#1{(\ref{#1})}

\def\rf#1{(\ref{#1})}

\def\rfs#1{Eq.~\rf{#1}}
\def\sign{{\rm sign}}

\begin{document}


\title{Nonequilibrium dynamics and thermodynamics of a degenerate
Fermi gas across a Feshbach resonance}


\author{A. V. Andreev}
\author{V. Gurarie}
\author{L. Radzihovsky}
\affiliation{Department of Physics, University of Colorado,
Boulder CO 80309}


\date{\today}

\begin{abstract}
  We consider a two-species degenerate Fermi gas coupled by a diatomic
  Feshbach resonance.
  We show that the resulting superfluid can exhibit a
  form of coherent BEC-to-BCS oscillations in response to a
  nonadiabatic change in the system's parameters, such as for example a
  sudden shift in the position of the Feshbach resonance. In the
  narrow resonance limit, the resulting soliton-like collisionless
  dynamics can be calculated analytically. In equilibrium the
  thermodynamics can be accurately computed across the full range of
  BCS-BEC crossover, with corrections controlled by the ratio of the
  resonance width to the Fermi energy.
\end{abstract}

\maketitle

Recent experimental advances to control interactions in trapped
degenerate gases by tuning (via magnetic field) through a Feshbach
resonance (FR), led to observation of long-sought after molecular
Bose-Einstein condensation in bosonic \cite{Donley} and fermionic
atomic gases \cite{Jin2004,Ketterle}.  Such tunability allows
studies of these systems in previously unexplored regimes, as
exemplified by recent JILA and MIT experiments
\cite{Jin2004,Ketterle}, that appear to observe paired fermionic
condensates along the crossover between the BCS regime of
weakly-paired, strongly overlapping Cooper pairs, and the BEC
regime of tightly bound, weakly-interacting diatomic molecules. An
even more exciting possibility, unavailable in other related
(e.g., superconductors and superfluids in condensed matter)
systems, is the nonadiabatic switching of system's parameters,
thereby allowing access to highly coherent and nonequilibrium
quantum states of matter. For bosonic $^{85}$Rb atoms, this was
recently spectacularly realized in experiments by Donley {\em et
al} \cite{Donley}. Using short magnetic field pulses that briefly
bring the system close to a nearby FR, they observed coherent
oscillations in the atomic condensate that can be interpreted as
Rabi oscillations between atomic and molecular condensates
\cite{Holland2002}.

Stimulated by these experimental advances, in this paper we study
a zero-temperature collisionless dynamics of a two-species
degenerate atomic Fermi gas near a FR, that can be tuned through
the Fermi sea. The goal is to understand the evolution of the
system following a nonadiabatic change in an externally-controlled
system's parameter, such as the detuning of the FR, $\omega_0$,
relative to the Fermi energy, $\epsilon_F$.

Our main result is the demonstration that such a system exhibits
an integrable soliton-like solutions, corresponding to collective
coherent oscillations of the Fermi gas between BCS- and BEC-like
paired superfluid states. This dynamics is a paired-fermions
analog of the atomic-to-molecular condensate Rabi oscillations
observed by Donley {\em et al} in trapped bosonic gases. In the
narrow Feshbach resonance limit we calculate the frequency and
amplitude of these oscillations and find their  analytic form. Our
work directly builds on the recent discovery by Barankov, Levitov,
and Spivak \cite{Barankov2003} of integrable dynamics in the BCS
model following a sudden change in the negative s-wave scattering
length.

Complementing our study of dynamics, we also analyze the
thermodynamics along the full range of the BCS-BEC crossover
\cite{Holland2001,Timmermans2001}.
We show that for a FR that is narrow compared to the Fermi energy,
and a background scattering length $a_{bg}$ that is short compared
to the interatomic spacing, $n^{-1/3}$, low-temperature
thermodynamics can be accurately computed analytically
\cite{largeGamma}.

Our starting point is the Hamiltonian \cite{trap}
\begin{eqnarray}
\label{eq:ham} H &=& \sum_{p, \sigma} \epsilon_p ~\hat
a^\dagger_{p \sigma} \hat a_{p \sigma} + \sum_p \left(\epsilon_0
+{\epsilon_p \over 2} \right) \hat b_p^\dagger \hat b_p\cr
&&+\sum_{p,q} ~{g \over \sqrt{V}} \left( \hat b_q ~ \hat
a^\dagger_{p+q \uparrow} \hat a^\dagger_{-p \downarrow} +
\hat b^\dagger_q ~ \hat a_{-p \downarrow} \hat a_{p+q \uparrow}
\right)
\end{eqnarray}
describing fermionic atoms, created by $\hat a^\dagger_{p\sigma}$
with momentum $p$, ``spin" $\sigma=\uparrow,\downarrow$, and
kinetic energy $\epsilon_p=p^2/2m$, that are coupled to diatomic
molecular (resonant) states created by $\hat b^\dagger_q$. The
position and the width (molecular lifetime) of the FR are
respectively controlled by the bare detuning energy $\epsilon_0$
and coupling $g$, the former easily experimentally tunable by a
magnetic field. We have neglected nonresonant atom-atom and
molecule-molecule interactions that we expect near a FR to be
subdominant to the resonant scattering.

In what follows we focus on molecules with zero center of mass
momentum, $\hat b_0$, neglecting molecules $\hat b_{q\neq0}$
excited above the molecular condensate. For equilibrium phenomena,
this approximation is justified at low temperatures and weak
interactions (small $g$), for which condensate depletion is small.
However, for nonequilibrium dynamics of interest to us, the
validity of this approximation is a more delicate issue, as one
might generically expect such bosonic finite momentum excitations
to be induced by a nonadiabatic shift in the FR. Nevertheless,
physically we expect that for an initially homogeneous condensate
\cite{trap} and weak interactions the dynamics will be dominated
by $b_0(t)$. However, on sufficiently long time scales the
$b_{q\neq 0}$ excitations and particle collisions should decohere
and damp out the collective BEC-BCS oscillations studied here,
allowing a slow relaxation to a new equilibrium state for the
shifted FR. Determining this relaxational dynamics is beyond the
scope of the present work. We thus replace $\hat b_q$ in
\rfs{eq:ham} by $\hat b_0 \delta_{{\bf
    q},0}\equiv \hat b \, \delta_{{\bf q},0}$.
Expected macroscopic occupation ($\langle \hat b^\dagger \hat
b\rangle \gg 1$) of the molecular level $\epsilon_0$ allows us to
neglect quantum molecular fluctuations and replace operators $\hat
b(t), \hat b^\dagger(t)$ by the corresponding c-numbers $b(t),
b^*(t)$.

We now look for time dependent fermionic operators in terms of
reference, time-independent fermions $\hat \alpha_{p \uparrow},
\hat \alpha_{-p\downarrow}$, related to $\hat a_{p,\sigma}(t)$
through the Bogoliubov amplitudes $u_p(t)$, $v_p(t)$ by $\hat
a_{p\uparrow} = u^*_p \, \hat \alpha_{p \uparrow} + v_p \, \hat
\alpha^\dagger_{-p\downarrow}$, $\hat a_{-p \downarrow} = u^*_p \,
\hat \alpha_{-p \downarrow} - v_p \, \hat \alpha^\dagger_{p
\uparrow}$ with a constraint $u^*_p u_p + v^*_p v_p=1$, that
ensures fermionic anticommutation relations. The Heisenberg
dynamics is then encoded in the equations of motion for $u_p(t)$,
$v_p(t)$, and $b(t)$
\begin{eqnarray}
\label{eq:uv} \hspace{-2cm} i \d_t u_p &=&  - \epsilon_p u_p+{g
b^*\over \sqrt{V}}  \, v_p, \, i
\d_t v_p = \epsilon_p v_p + {g b\over \sqrt{V}}  \, u_p, \\
\label{eq:b} i \d_t b &=& \epsilon_0 b + {g \over \sqrt{V}} \sum_p
u^*_p v_p.
\end{eqnarray}

The dynamical evolution equations (\ref{eq:uv}) and (\ref{eq:b})
preserve pair correlations of fermions. Accordingly, we choose the
initial state to be of the BCS type, $\left|\Omega(0)
\right\rangle = \prod_p \left( u_p(0) + v_p(0)~ a^\dagger_{p
\uparrow} a^\dagger_{-p \downarrow} \right) \left|  0
\right\rangle$. As a result of the time evolution the fermion wave
function preserves the same form with time dependent factors
$u_p(t)$ and $v_p(t)$.

Following Ref.~\cite{Barankov2003} we utilize Anderson's spin
analogy for the BCS problem \cite{Anderson}, and look for a
vector-representation of these equations in terms of the variables
\begin{equation}
S_p = S^x_p + i S^y_p = 2 v^*_p u_p, \ \ \ S^z_p=v_p v^*_p - u_p u^*_p,
\end{equation}
that satisfy $|{\bf S}_p|^2=S^*_p S_p + S^z_p S^z_p =1$. Equations
(\ref{eq:uv}) become
\begin{eqnarray}
\label{eq:vo} \d_t S_p &=& 2 i \epsilon_p \, S_p - {2 i g \over
\sqrt{V}} \, b^* S^z_p, \ \d_t  S^*_p = - 2 i \epsilon_p  \, S^*_p
+ {2 i g\over \sqrt{V}}  b \,S^z_p, \cr \d_t S^z_p &=& {i g \over
\sqrt{V}}  \left( b^* S^*_p - b \, S_p \right),
\end{eqnarray}
and have a form of Bloch equations for a spin precessing in an
effective ($p$-dependent) field $(-{2 g \over \sqrt{V}}
\,\text{Re}\,b, {2 g\over \sqrt{V}} \,\text{Im}\,b, - 2
\epsilon_p)$, whose azimuthal dynamics is in turn
self-consistently determined by molecular ($b(t)$) evolution,
\rfs{eq:b}. It is now straightforward to check that these
equations can be solved by an ansatz (for the BCS limit, in the
absence of molecules suggested in Ref.~\cite{Barankov2003})
\begin{eqnarray}
\label{eq:ansatz} S_p(t) &=& e^{i \omega t} \left( - (2 \epsilon_p
- \omega) C_p \Omega(t) + i  C_p \dot \Omega(t) \right), \cr
S^z_p(t) &=&  C_p \Omega^2(t) - D_p, \ {g\, b(t)\over \sqrt{V}} =
{\Omega(t)} \, e^{- i \omega t},
\end{eqnarray}
provided that the real function $\Omega(t)$ satisfies
\begin{equation}
\label{eq:omega} \dot \Omega^2 + \left( \Omega^2 - \Delta_-^2
\right) \left( \Omega^2 - \Delta_+^2 \right) =0,
\end{equation}
and that $C_p$ and $D_p$ satisfy
\begin{equation}
\label{eq:CD} {D_p^2 - 1 \over C_p^2} = \Delta_-^2 \Delta_+^2, \ 2
{D_p \over C_p} = \left( 2 \epsilon_p - \omega \right)^2 +
\Delta_-^2 + \Delta_+^2.
\end{equation}
Here $\Delta_+$, $\Delta_-$, and $\omega$ are parameters
characterizing the periodic instanton-like solution $\Omega(t)$
expressible in terms an elliptic integral. It follows from
\rfs{eq:omega} that $b(t)/\sqrt{V}$ oscillates between the values
$\Delta_-/g$ and $\Delta_+/g$ (we choose $\Delta_-<\Delta_+$) with
a period of oscillations given by
\begin{equation}
\label{eq:period} T={2 \over \Delta_+} K\left(1-{\Delta_-^2 \over
\Delta_+^2} \right),
\end{equation}
where $K(m)$ is the complete elliptic integral of the first kind.

A direct substitution of this ansatz into \rfs{eq:b}, shows that the
solution, \rfs{eq:ansatz} is compatible with the evolution of $b(t)$
provided that the following two conditions are satisfied:
\begin{eqnarray}
\epsilon_0 - \omega &=& {g^2 \over 2} \int {d^3 p
\over (2 \pi)^3} \left( 2 \epsilon_p - \omega \right)
C_p, \label{eq:gap}\\
1 &=& -{g^2 \over 2} \int {d^3 p \over (2 \pi)^3} \, C_p.
\label{eq:conserve}
\end{eqnarray}
Eqs.~\rf{eq:gap}, \rf{eq:conserve} determine $\omega$ and
$\Delta_+$ in terms of the experimentally controlled (initial
conditions) parameters $\epsilon_0$ and $\Delta_-$.  As we will
see below, the first of these equations is a nonequlibrium
generalization of the BCS gap equation. The second one simply
reflects the conservation of the total particle number, ${d N\over
dt}=0$, with
\begin{equation}
\label{eq:N} N = 2 b^* b + V \int {d^3 p \over (2 \pi)^3}
\left(S^z_p+1 \right).
\end{equation}

We note that Eqs.~\rf{eq:CD} only determine $C_p$ and $D_p$ up to
a $p$-dependent sign, that one would expect to be fixed by the
initial fermion momentum distribution, encoded in
$|\Omega(0)\rangle$. With the exception of a filled Fermi sea
initial condition, $\Omega(0)=\Delta_-=0$, the solution encoded in
$C_p$ and $D_p$, \rfs{eq:CD}, does {\em not} correspond to initial
conditions ($u_p(0)$ and $v_p(0)$) characteristic of a ground
state of the Hamiltonian \rf{eq:ham} for any value of bare
detuning $\epsilon_0$. Nevertheless we fix the sign of $C_p$ so as
to most closely match $n_p=\oh (S^z_p+1)$ to the initial fermion
momentum distribution.  For a large positive detuning this
corresponds to a Fermi-Dirac step function with discontinuity at
the chemical potential $\mu$, that separates the hole-like and
particle-like states.  Combining this criterion with
Eqs.(\ref{eq:CD}), we find
\begin{equation}
C_p = {2 ~{\rm sign} (\epsilon_p - \mu) \over \sqrt{ \left[
(2\epsilon_p - \omega)^2 + {\Delta_-}^2 + {\Delta_+}^2 \right]^2 - 4
\Delta_-^2 \Delta_+^2}},
\end{equation}
where, as in equilibrium problem, $\mu$ is implicitly determined
by the conserved total particle number $N$, \rfs{eq:N}, that
reduces to $N= V \int {d^3 p \over (2 \pi)^3} \,
\left(1-D_p\right)$.

Equation (\ref{eq:gap}) involves a linearly-divergent
integral~\cite{Gorkov1961}, arising from an unphysical aspect of
the model \rfs{eq:ham}, that the atomic modes with an arbitrarily
large energy interact with the molecular ones with equal strength
$g$.  In a more realistic model, the momentum dependence of the
coupling $g$ would cutoff this divergence.  As usual, our
ignorance of this high energy physics can be buried in a (uv
cutoff-dependent) relation between the parameter $\epsilon_0$
appearing in the Hamiltonian and the position of the physical
Feshbach resonance, $\omega_0$.

To see this, we calculate the two-atom scattering amplitude within
the model Hamiltonian \rfs{eq:ham}. It is completely determined by
the self-energy of the molecules, given by
\begin{eqnarray}
\Sigma(E) &=& \int {d \omega \over 2\pi} {d^3 p \over (2 \pi)^3}
{i g^2 \over \left( E - \omega - \epsilon_p + i 0^+ \right) \left(
\omega - \epsilon_p + i 0^+\right)}\cr &=& -i { g^2 \over 4 \pi}
m^{3 \over 2} \sqrt{ E}-g^2 \int {d^3 p \over (2 \pi)^3} {m \over
p^2},
\end{eqnarray}
that diverges in exactly the same way as the integral in
\rfs{eq:gap}. Since $\Sigma(E)$ enters the retarded molecular
propagator $G_b(E)= \left(E -
  \epsilon_0 -\Sigma(E) \right)^{-1}$ in combination with $\epsilon_0$,
we can trade in $\epsilon_0$ for the physical (``renormalized") FR
detuning $\omega_0$ according to
$
 \omega_0 = \epsilon_0 - g^2 \int {d^3 p \over (2\pi)^3} {m \over p^2}.
$ This leads to the two-atom scattering amplitude $ f = -{1 \over
\sqrt{m}} {\gamma \over  E - \omega_0 + i \gamma \sqrt{E}},$ that
is consistent with the generic form based on unitarity \cite{LL}
and is identical to that of the Fano-Anderson model (equivalent to
our model at the two-body scattering level) \cite{Fano1961}. We
see that the scattering phase changes by $\pi$ as the energy
changes from below to above the physical (``renormalized'') FR,
$\omega_0$, with finite width for positive detuning, $\omega_0>0$,
controlled by $\gamma = g^2 m^{3/2}/4\pi$.  In contrast for
negative detuning, $\omega_0<0$, the scattering amplitude has a
real pole at negative energy (a bound state in the open channel),
that corresponds to a real molecular state \cite{claims}.

With above renormalization of  detuning, the nonequilibrium gap
equation \rf{eq:gap} becomes
\begin{eqnarray}
\label{eq:gapRenorm} \omega_0 - \omega &=& {g^2 \over 2} \int {d^3 p
\over (2 \pi)^3} \left[ \left( 2 \epsilon_p - \omega \right) C_p -
{2 m \over p^2} \right],
\end{eqnarray}
where all the integrals are now convergent. This together with the
atom conservation condition, \rf{eq:conserve}, the total atom
number \rfs{eq:N} (that determines $\mu$), the detuning
$\omega_0$, and the initial molecular density $\Delta_-$, allows
us to determine the condensate frequency $\omega$, and the
molecular density maximum $\Delta_+$, that controls the period of
oscillations, in accordance with \rfs{eq:period}. Although the
complete solution requires a numerical evaluation of the
integrals, here we will focus on two analytically tractable
regimes: (i) $\Delta_+-\Delta_- \ll \Delta_+$ that corresponds to
small amplitude oscillations of the condensate about the BCS-BEC
ground state, and (ii) $\Delta_- \ll \Delta_+$, that corresponds
to the evolution of the filled Fermi sea, following a large
downward shift in detuning.

Let us first consider the regime of small oscillations,
$\Delta_+-\Delta_- \ll \Delta_+$ about a BCS-BEC ground state.
Limiting our analysis to $\omega>0$, we find that Eq.~\rf{eq:conserve}
constrains $\omega$ to be close to $2\mu$.  This together with the
condition \rfs{eq:gapRenorm} gives (at $\Delta_-=\Delta_+=\Delta$, and
$\mu=\omega/2$)
\begin{eqnarray}
\label{eq:bcsbec} \omega_0 - 2 \mu = {g^2 \over 2} \int {d^3 p
\over (2 \pi)^3 } \left[ { 1\over \sqrt{\left( \epsilon_p - {\mu }
\right)^2 + \Delta^2 }} - {1 \over \epsilon_p} \right].
\end{eqnarray}
This coincides with the BCS-BEC gap equation, that can be derived
in the equilibrium treatment of this problem
\cite{us_unpublished}. It relates the condensate density
$\Delta/g$ to the Feshbach resonance $\omega_0$, with the chemical
potential $\mu$ determined by the total atom number equation,
Eq.~\rf{eq:N}. Simple analysis of these equations shows that for a
large positive detuning, $\omega_0\gg \epsilon_F$, molecules are
strongly suppressed, leading the chemical potential to ``stick''
to $\epsilon_F$, and to a conventional atomic BCS ground state,
with $\Delta(\omega_0)\approx 8 e^{-2}\epsilon_F
e^{-(\omega_0-2\epsilon_F)/g^2\nu(\epsilon_F)}$
($\nu(\epsilon)=m^{3 \over 2} \epsilon^{1 \over 2}/\sqrt{2} \pi^2$
is the atomic density of states).  In this far off-resonance BCS
regime the accuracy of the mean-field treatment of the equilibrium
problem is controlled by the ratio of the width of the FR to Fermi
energy, namely by the dimensionless parameter
$\gamma^2/\epsilon_F$ \cite{largeGamma}.

As the detuning $\omega_0$ is lowered toward and below $\epsilon_F$,
the chemical potential begins to track the detuning,
$\mu(\omega_0)\approx \omega_0/2 - {\cal O}[g^2\nu(\omega_0/2)]$, with
atoms from states between $\epsilon_F$ and $\mu(\omega_0)$ converting
into Bose-condensed molecules. The density of these tightly bound
molecules, that coexist with BCS's Fermi sea determine the gap, which
displays a rounded mean-field behavior $\Delta_{\text
  equil}(\omega_0)\approx\sqrt{{2^{5/2}\over
    3\pi}\gamma(\epsilon_F^{3/2}-\mu(\omega_0)^{3/2})}$. In the
$g\rightarrow 0$ limit, $\omega_0$ crossing of $\epsilon_F$ is a
genuine quantum transition, with an upper-critical dimension of
$d_{uc}=2$, and is therefore mean-field in 3d \cite{Uzunov}. A
finite atom-molecule coupling $g$ rounds the transition into a
smooth crossover near $\epsilon_F$, that for small $g$ (i.e.,
narrow FR, $\gamma^2\ll\epsilon_F$) is therefore also accurately
described by the mean-field theory summarized by
Eqs.(\ref{eq:N},\ref{eq:bcsbec}). Clearly, no additional anomalies
appear when the 2-body FR ($\omega_0=0$) is crossed, since by that
point nearly all atoms are bound up into well-ordered
Bose-condensed molecules. One interesting observation is that in
this $\mu<0$ BEC regime the remaining dilute fermionic atoms are
paired but are nondegenerate, and therefore realize a
``strongly-coupled'' BCS state \cite{Read2000}.  This picture of
the ground state (that can be easily extended to a finite
temperature\cite{us_unpublished}) is qualitatively consistent with
recent observations by Regal, et al.\cite{Jin2004}, which find
that molecules appear at about 0.5 Gauss above the experimentally
determined 2-body FR (in our interpretation corresponding to
$\omega_0 \approx 2\epsilon_F$).

The nonequilibrium solution, \rfs{eq:ansatz} then describes small
oscillations about this equilibrium BCS-BEC state, with the period
$T=\pi/\Delta$ given by \rfs{eq:period}. Because $\omega\approx
2\mu$ it is easy to see that for a sufficiently small oscillation
amplitude, the atomic momentum distribution function $n_p(t)=\oh
(S^z_p(t)+1)$ only changes near the Fermi surface,
$\epsilon_p=\mu$.

The other interesting limit, $\Delta_- \rightarrow 0$,
$\Delta_+=\Delta$ describes oscillations between a Fermi sea (a
normal ``metal'', with $\mu=\epsilon_F$) of atoms and bosonic
molecules, following a nonadiabatic shift of detuning from
$\infty$ down to $\omega_0$.  In this regime,
\rfs{eq:conserve},\rf{eq:gapRenorm} reduce to
\begin{eqnarray}
\label{eq:shift_detuning}
\omega_0 - \omega &=& {g^2 \over 2} \int {d^3 p \over (2 \pi)^3}
\left[ {\left(  \epsilon_p - {\omega  \over 2} \right) \sign\left(
\epsilon_p - {\epsilon_F} \right) \over \left(  \epsilon_p - {\omega
\over 2} \right)^2 + {\Delta^2 \over 4} } - {1 \over \epsilon_p}
\right], \cr
1 &=& - {g^2 \over 4} \int {d^3 p \over (2 \pi)^3} {
\sign\left( \epsilon_p - {\epsilon_F} \right) \over \left(  \epsilon_p -
{\omega \over 2} \right)^2 + {\Delta^2 \over 4} }.
\end{eqnarray}

Simple integration of the atom-number conservation (second) equation
above gives
\begin{equation}
\Delta\approx
g^2\nu(\omega/2)\tan^{-1}[(2\epsilon_F-\omega)/\Delta].
\label{eq:conserveSolve}
\end{equation}
The behavior then depends qualitatively on whether the detuning
$\omega_0$ is shifted to the BCS ($\omega_0/2 >> \epsilon_F$) or
to the BCS-BEC ($0<\omega_0/2 <\epsilon_F$) regime, i.e., above or
below the Fermi surface. In the former case \rfs{eq:conserveSolve}
reduces to $2\epsilon_F-\omega\approx
\Delta^2/g^2\nu(\omega/2)\ll\Delta$, with an exponentially
suppressed BCS-like gap $\Delta \propto e^{-{\omega_0 -
2\epsilon_F \over g^2\nu(\epsilon_F)}}$ given by the gap (first)
equation in \rfs{eq:shift_detuning}. Hence in this regime, for
narrow FR, $\omega\approx 2\epsilon_F$ and oscillations are
confined to the vicinity of the Fermi surface.

\begin{figure}[bth]
\centering \setlength{\unitlength}{1mm}
\begin{picture}(40,31)(0,0)
\put(-35,0){\begin{picture}(20,20)(0,0)
\includegraphics{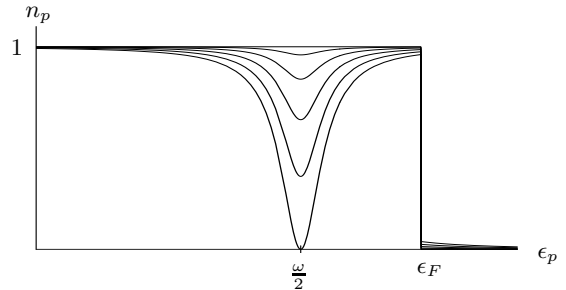}
\end{picture}}
\put(-10,32) {$n_p$} \put(-12,27) {$1$} \put(58,0) {$\epsilon_p$}
\put(25,-3){${\omega\over 2}$} \put(42,-2) {$\epsilon_F$}
\end{picture}
\caption{Atomic momentum distribution function $n_p$ displaying
oscillations
  between a Fermi-sea and Bose-condensed molecules. The atomic
  depletion in $n_p$ of width $\Delta\approx\gamma\sqrt{\omega_0}$
  appears at the molecular energy determined by the detuning
  $\omega_0$.}
\label{BEC-FLoscillate}
\end{figure}

In contrast, for the detuning shift {\em below} the Fermi surface,
$0 < \omega_0/2 < \epsilon_F$, it is now \rfs{eq:conserveSolve}
that determines the molecular density. It gives $\Delta\approx
g^2\nu(\omega/2)/2 = \gamma\sqrt{\omega} \times {\cal O}(1)$,
while the gap equation gives $\omega_0-\omega = g^2\nu(\epsilon_F)
\times {\cal O}(1) \ll \omega_0$, enforcing the molecular energy
$\omega$ to stick to the detuning $\omega_0$.  We note, that,
although $\Delta\approx\gamma\sqrt{\omega_0}$ is much larger
(scales as $g^2$ rather than exponentially suppressed in $1/g^2$)
than for the detuning into the BCS regime, it is much smaller than
the equilibrium $\Delta_{\text equil}(\omega_0)$, found as the
solution to the equilibrium gap equation \rfs{eq:bcsbec} in this
regime. This suppression of the condensate oscillations is due to
energy conservation between the atoms in the Fermi sea and the
molecules.  For a narrow FR resonance, $\gamma\sqrt{\omega_0} \ll
\omega_0 < \epsilon_F$ it is only a small fraction of atoms in the
Fermi sea that are in the immediate vicinity (set by the resonance
width $\gamma\sqrt{\omega_0}$) of $\omega_0$, that can resonantly
bind into molecules.  The resulting resonant atom-molecule
interconversion leads to a narrow oscillatory depletion of the
Fermi sea, illustrated in Fig.\ref{BEC-FLoscillate}, with a period
of oscillations that diverges as $T \propto {2 \over \Delta_+}
\log \left( \Delta_+ \over \Delta_- \right)$, in the limit of the
``normal'' Fermi sea ($\Delta_- \rightarrow 0$) as the initial
condition.

Analysis of the gap and atom-conservation equations shows that the
amplitude of the atom-molecule oscillations vanishes as the
molecular energy $\omega\approx \omega_0$ approaches zero. This is
again enforced by the energy conservation that in the absence of
other degrees of freedom (e.g., molecules above a condensate,
$b_{q\neq0}$) forbids conversion of molecules at negative energy
$\omega_0$ into atoms at positive energy $\epsilon_p$. We leave
analysis of the dynamics that incorporates these additional
degrees of freedom to future research.

We acknowledge support by the NSF DMR-0321848 (L.R.), DMR-9984002
(A.A.), and the Packard Foundation (A.A., L.R.), and thank J.
Bohn, C. Greene, and D. Jin for discussions.

\vspace{-0.5cm}

\bibliography{submit}

\begin{thebibliography}{17}
\bibitem[{tra()}]{trap}
\bibinfo{note}{For simplicity, we focus on a homogenous system of volume $V$,
  leaving extension to the case of the trap for future work.}
\bibitem[{cla()}]{claims}
\bibinfo{note}{We note here that although at vanishing $E$ the scattering
  amplitude diverges as $1/\omega_0$, for a narrow resonance,
  $\gamma\ll\sqrt{\epsilon_F}$, a dilute system is still weakly interacting and
  therefore amenable to a controlled perturbative treatment (in contrast to
  claims in the literature).}
\end{thebibliography}

\end{document}